\documentclass[a4paper,12pt]{article}
\usepackage[utf8]{inputenc}
\usepackage[english]{babel}
\usepackage{csquotes}
\usepackage{authblk}
\usepackage{graphicx}
\usepackage{mathptmx}
\usepackage[singlespacing]{setspace}
\usepackage[headheight=1in,margin=1in]{geometry}
\usepackage{fancyhdr}
\usepackage{lipsum}
\usepackage{biblatex}
\usepackage{authblk}
\def\@maketitle{%
  \newpage

  \begin{center}%
    {\LARGE \@title \par}%
    \vskip 1em
    {\Small \@author \par}%
  \end{center}%
  \par
  \vskip 0.1em}

\graphicspath{{images/}}

\title{Who Provides the Largest Megaphone? The Role of Google News in Promoting Russian State-Affiliated News Sources}

\date{}
\author[1]{Keeley Erhardt}
\author[2]{Saurabh Khanna}
\affil[1]{Massachusetts Institute of Technology, Cambridge, MA}
\affil[2]{Stanford University, Stanford, CA}

\begin{document}

\maketitle

\thispagestyle{fancy}

\begin{center}
\textit{Keywords: Ranking algorithms, search engines, online news, state media, propaganda}
\newline
\end{center}

\subsection*{Extended Abstract}

\subsubsection*{Introduction}

The Internet has not only digitized but also democratized information access across the globe. This gradual but path-breaking move to online information propagation has resulted in search engines playing an increasingly prominent role in shaping access to human knowledge. When an Internet user enters a query, the search engine sorts through the hundreds of billions of possible webpages to determine what to show. Google dominates the search engine market, with Google Search surpassing 80\% market share globally every year of the last decade. Only in Russia and China do Google competitors claim more market share, with $\approx60\%$ of Internet users in Russia preferring Yandex (compared to 40\% in favor of Google) and more than 80\% of China’s Internet users accessing Baidu as of 2022.\footnote{https://www.statista.com/statistics/216573/worldwide-market-share-of-search-engine} Notwithstanding this long-standing regional variation in Internet search providers, there is limited research showing how these providers compare in terms of propagating state-sponsored information. 

Our study fills this research gap by focusing on Russian cyberspace and examining how Google and Yandex’s search algorithms rank content from Russian state-controlled media (hereon, RSM) outlets. This question is timely and of practical interest given widespread reports indicating that RSM outlets have actively engaged in promoting Kremlin propaganda in the lead-up to, and in the aftermath of, the Russian invasion of Ukraine in February 2022.

\subsubsection*{Methods}

 We consider six RSM outlets in our study: RIA Novosti, Lenta, Gazeta, TASS, RT, and Izvestia. Our dataset consists of Google Search and Yandex Search queries and results from Russia from 2021-12-01 to 2022-03-01. We collected the search data by first identifying the top daily trending queries for \url{google.ru} from Google Trends. For Google Search, we then sent the trending queries to a privately hosted instance of Searx, an open-source meta-search engine, from an IP address in Russia, and logged all returned search results. Some queries were blocked by Searx. For these, we instead leveraged the Google Search Engine Results Page (SERP) API provided by DataForSEO. For Yandex Search, we collected search results for each of the same set of queries using SerpWow's Yandex SERP API. Though the queries trending on Google are not necessarily the same as those trending on Yandex, this data collection method ensured that the same set of queries was used to collect search results across the two search engines.

\subsubsection*{Results}

As seen in Figure 1, we find that across all trending queries in Russia, an average of 7.74\% of the articles on the first page of Google search results belonged to RSM outlets. For Yandex, this proportion is lower at 3.73\% of articles but steadily increases over the 90-day analysis window. These numbers suggest that the average Internet user in Russia was exposed to a sizable volume of RSM content whether using Google Search or Yandex Search. Further, they indicate that Google's algorithms ranked RSM content more favorably than Yandex's by a factor of two.

A possible explanation for the difference is that Google's search algorithms assign a higher rank to news media sources in general (RSM or otherwise), compared to Yandex. To test this theory, we defined a set of control media outlets consisting of two popular international news sources in Russia, the BBC (\url{bbc.com}) and Forbes (\url{forbes.ru}), and three independent, domestic media outlets, TV Rain (\url{tvrain.com}), Novaya Gazeta (\url{novayagazeta.ru}), and Interfax (\url{interfax.ru}).
We then analyzed how often webpages from these news outlets appeared as results for trending queries. We found that content from the control outlets appeared more frequently on Google Search than Yandex Search by a factor of 3.45, indicating that Google's algorithms do appear to assign a higher weight to news media sources (whether implicitly or explicitly). Still, both search engines return substantively more results from RSM sources compared to the control -- a fourfold increase for Google and sevenfold increase for Yandex.

\subsubsection*{Discussion}

We recognize three limitations in our study. First, Google Trends is an imperfect representation of search queries, as previously described in the literature~\cite{rovetta2021reliability}. Notably, it does not provide search volumes and instead aggregates queries and compares their relative variations across time. We believe this limitation is acceptable as we are most interested in the webpages that appear as search results, rather than search volumes. Second, we collected Google search results using Searx and DataForSEO, and Yandex results using SerpWow. This was due to Searx and DataForSEO imposing restrictions on access to their services from IP addresses in Russia after the invasion began in February 2022.
Third, our analysis covers only a subset of RSM outlets, and our control group is limited to five independent media organizations. Still, we believe that given the prevalence of these outlets in Russia, the results are broadly representative.


This work contributes timely and important insight into the role of two of the largest search engines in sorting, and thereby, implicitly recommending, content from RSM sources amidst an ongoing armed conflict. Our results contradict the common assumption that Yandex is the primary driver of an online information environment over-represented by RSM content in Russia. Instead, while Yandex does appear to rank RSM content more highly than Google relative to the rate at which it returns results from news media sources, both search engines frequently surface content from Russia state-controlled sources and, in fact, Google returns RSM at a higher absolute rate. This finding has serious implications for our understanding of the dissemination and visibility of state-promoted content online.

\subsection*{Acknowledgements}
Research was sponsored by the United States Air Force Research Laboratory and the Department of the Air Force Artificial Intelligence Accelerator and was accomplished under Cooperative Agreement Number FA8750-19-2-1000. The views and conclusions contained in this document are those of the authors and should not be interpreted as representing the official policies, either expressed or implied, of the Department of the Air Force or the U.S. Government. The U.S. Government is authorized to reproduce and distribute reprints for Government purposes notwithstanding any copyright notation herein.

{\renewcommand{\markboth}[2]{}\printbibliography}




\begin{figure}[htb]
  \centering
  \includegraphics[width=\linewidth]{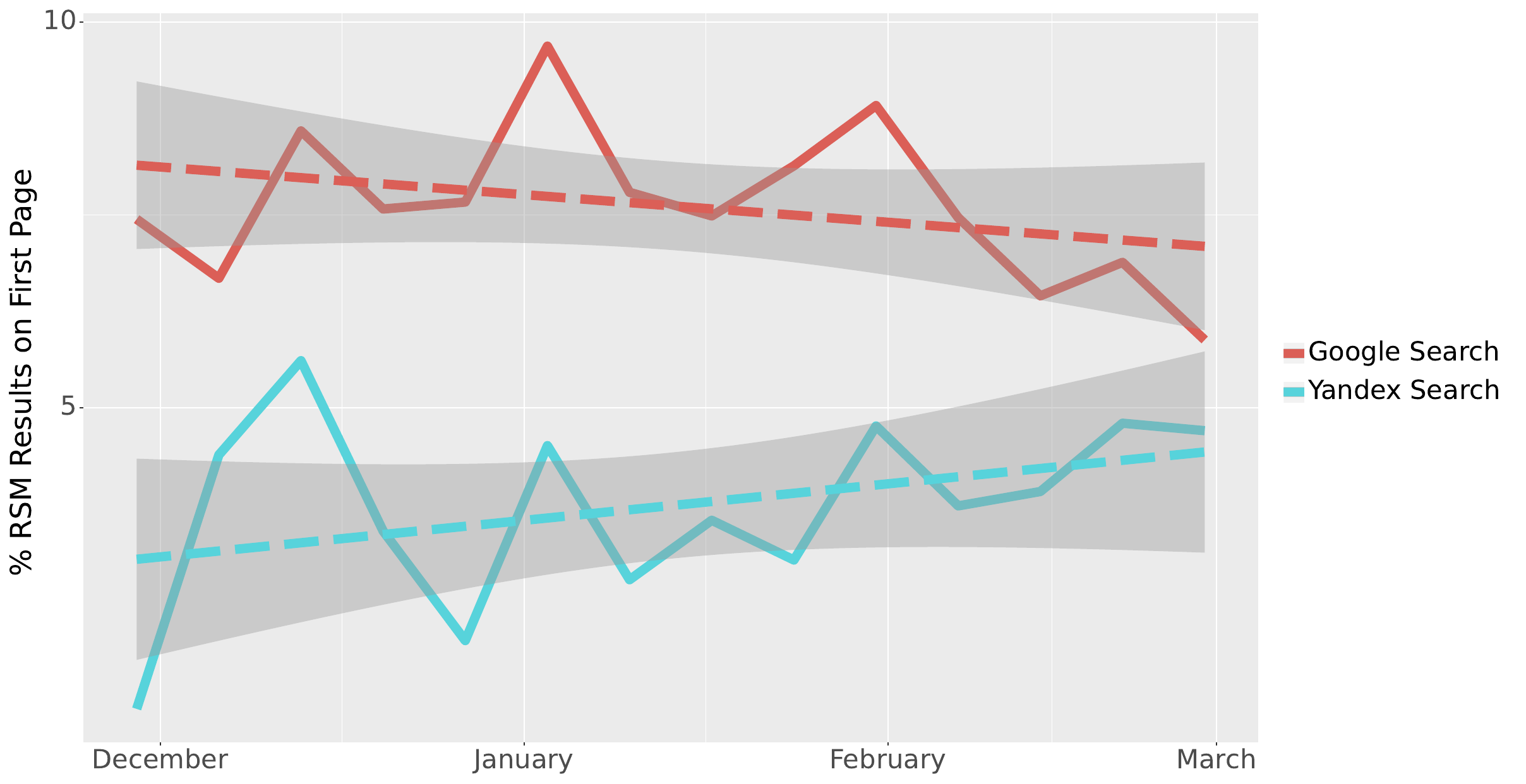}
  \caption{Percentage of Russia state-controlled news media results on the first page of search results across daily trending queries on Google Search and Yandex Search. The dotted lines represents smoothed conditional means and the gray region indicates the corresponding 95\% confidence intervals.}
  \label{fig:rsmPercentage}
\end{figure}

\begin{figure}[htb]
  \centering
  \includegraphics[width=\linewidth]{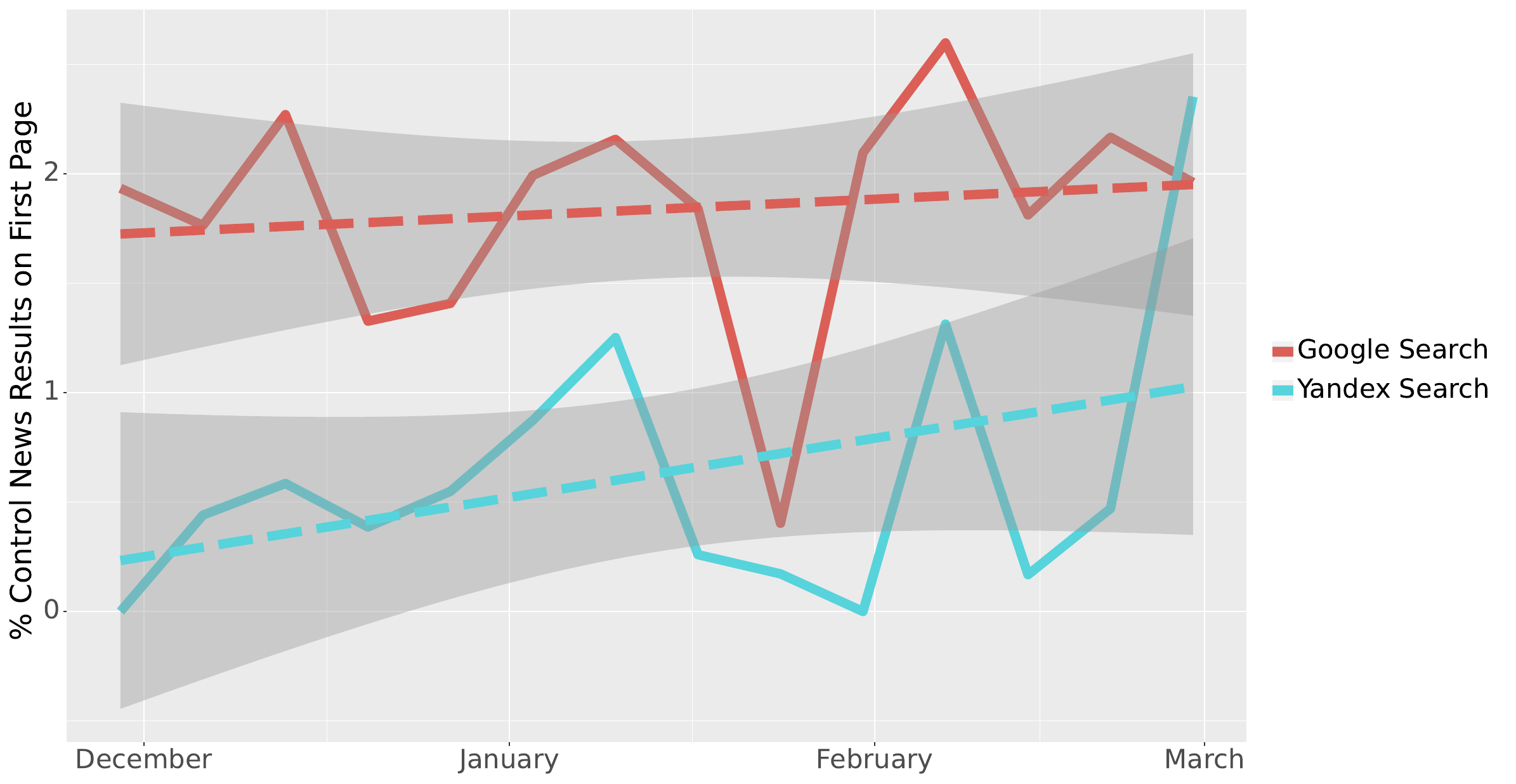}
  \caption{Percentage of Google Search and Yandex Search results from a sample of international news outlets and independent, Russian media outlets on the first page of search results across daily trending queries from Google Trends. The dotted lines represents smoothed conditional means and the gray region indicates the corresponding 95\% confidence intervals.}
  \label{fig:controlPercentage}
\end{figure}


\end{document}